\documentclass[9pt, conference]{IEEEtran}
\IEEEoverridecommandlockouts
\usepackage{amsmath,amssymb,amsfonts}
\usepackage{algorithmic}
\usepackage{graphicx}
\usepackage{booktabs}
\usepackage{multirow}
\usepackage{makecell}
\usepackage{textcomp}
\usepackage{enumitem}
\usepackage{xcolor}
\usepackage[hidelinks]{hyperref}
\usepackage{easyReview}
\usepackage{balance}
\usepackage{acronym}
\usepackage[style=ieee, doi=false, isbn=false, url=false, eprint=false, related=false, maxnames=6, minnames=1, date=year]{biblatex}
\usepackage{dblfloatfix}    
\AtEveryBibitem{\clearlist{language}}
\AtEveryBibitem{\clearfield{note}}
\AtEveryBibitem{\clearfield{address}}
\AtEveryBibitem{\clearlist{address}}
\AtEveryBibitem{\clearlist{location}}
\AtEveryBibitem{\clearfield{location}}
\AtEveryBibitem{\clearfield{publisher}}
\AtEveryBibitem{\clearlist{publisher}}
\AtEveryBibitem{\clearlist{pages}}
\AtEveryBibitem{\clearfield{pages}}
\AtEveryBibitem{\clearfield{pages}}
\usepackage{xurl}
\usepackage{forloop}

\def\BibTeX{{\rm B\kern-.05em{\sc i\kern-.025em b}\kern-.08em
    T\kern-.1667em\lower.7ex\hbox{E}\kern-.125emX}}

\newcommand{\ua}{$\uparrow$}
\newcommand{\da}{$\downarrow$}
\newcounter{loopcntr}

\usetikzlibrary{arrows.meta}

\makeatletter

\pgfdeclarearrow{
  name = ipe _linear,
  defaults = {
    length = +1bp,
    width  = +.666bp,
    line width = +0pt 1,
  },
  setup code = {
    \pgfarrowssetbackend{0pt}
    \pgfarrowssetvisualbackend{
      \pgfarrowlength\advance\pgf@x by-.5\pgfarrowlinewidth}
    \pgfarrowssetlineend{\pgfarrowlength}
    \ifpgfarrowreversed
      \pgfarrowssetlineend{\pgfarrowlength\advance\pgf@x by-.5\pgfarrowlinewidth}
    \fi
    \pgfarrowssettipend{\pgfarrowlength}
    \pgfarrowshullpoint{\pgfarrowlength}{0pt}
    \pgfarrowsupperhullpoint{0pt}{.5\pgfarrowwidth}
    \pgfarrowssavethe\pgfarrowlinewidth
    \pgfarrowssavethe\pgfarrowlength
    \pgfarrowssavethe\pgfarrowwidth
  },
  drawing code = {
    \pgfsetdash{}{+0pt}
    \ifdim\pgfarrowlinewidth=\pgflinewidth\else\pgfsetlinewidth{+\pgfarrowlinewidth}\fi
    \pgfpathmoveto{\pgfqpoint{0pt}{.5\pgfarrowwidth}}
    \pgfpathlineto{\pgfqpoint{\pgfarrowlength}{0pt}}
    \pgfpathlineto{\pgfqpoint{0pt}{-.5\pgfarrowwidth}}
    \pgfusepathqstroke
  },
  parameters = {
    \the\pgfarrowlinewidth,%
    \the\pgfarrowlength,%
    \the\pgfarrowwidth,%
  },
}

\pgfdeclarearrow{
  name = ipe _pointed,
  defaults = {
    length = +1bp,
    width  = +.666bp,
    inset  = +.2bp,
    line width = +0pt 1,
  },
  setup code = {
    \pgfarrowssetbackend{0pt}
    \pgfarrowssetvisualbackend{\pgfarrowinset}
    \pgfarrowssetlineend{\pgfarrowinset}
    \ifpgfarrowreversed
      \pgfarrowssetlineend{\pgfarrowlength}
    \fi
    \pgfarrowssettipend{\pgfarrowlength}
    \pgfarrowshullpoint{\pgfarrowlength}{0pt}
    \pgfarrowsupperhullpoint{0pt}{.5\pgfarrowwidth}
    \pgfarrowshullpoint{\pgfarrowinset}{0pt}
    \pgfarrowssavethe\pgfarrowinset
    \pgfarrowssavethe\pgfarrowlinewidth
    \pgfarrowssavethe\pgfarrowlength
    \pgfarrowssavethe\pgfarrowwidth
  },
  drawing code = {
    \pgfsetdash{}{+0pt}
    \ifdim\pgfarrowlinewidth=\pgflinewidth\else\pgfsetlinewidth{+\pgfarrowlinewidth}\fi
    \pgfpathmoveto{\pgfqpoint{\pgfarrowlength}{0pt}}
    \pgfpathlineto{\pgfqpoint{0pt}{.5\pgfarrowwidth}}
    \pgfpathlineto{\pgfqpoint{\pgfarrowinset}{0pt}}
    \pgfpathlineto{\pgfqpoint{0pt}{-.5\pgfarrowwidth}}
    \pgfpathclose
    \ifpgfarrowopen
      \pgfusepathqstroke
    \else
      \ifdim\pgfarrowlinewidth>0pt\pgfusepathqfillstroke\else\pgfusepathqfill\fi
    \fi
  },
  parameters = {
    \the\pgfarrowlinewidth,%
    \the\pgfarrowlength,%
    \the\pgfarrowwidth,%
    \the\pgfarrowinset,%
    \ifpgfarrowopen o\fi%
  },
}

\newdimen\ipeminipagewidth

\tikzstyle{ipe import} = [
  x=1bp, y=1bp,
  ipe node stretch/.store in=\ipenodestretch,
  ipe stretch normal/.style={ipe node stretch=1},
  ipe stretch normal,
  ipe node/.style={
    anchor=base west, inner sep=0, outer sep=0, scale=\ipenodestretch
  },
  ipe mark scale/.store in=\ipemarkscale,
  ipe mark tiny/.style={ipe mark scale=1.1},
  ipe mark small/.style={ipe mark scale=2},
  ipe mark normal/.style={ipe mark scale=3},
  ipe mark large/.style={ipe mark scale=5},
  ipe mark normal, 
  ipe circle/.pic={
    \draw[line width=0.2*\ipemarkscale]
      (0,0) circle[radius=0.5*\ipemarkscale];
    \coordinate () at (0,0);
  },
  ipe disk/.pic={
    \fill (0,0) circle[radius=0.6*\ipemarkscale];
    \coordinate () at (0,0);
  },
  ipe fdisk/.pic={
    \filldraw[line width=0.2*\ipemarkscale]
      (0,0) circle[radius=0.5*\ipemarkscale];
    \coordinate () at (0,0);
  },
  ipe box/.pic={
    \draw[line width=0.2*\ipemarkscale, line join=miter]
      (-.5*\ipemarkscale,-.5*\ipemarkscale) rectangle
      ( .5*\ipemarkscale, .5*\ipemarkscale);
    \coordinate () at (0,0);
  },
  ipe square/.pic={
    \fill
      (-.6*\ipemarkscale,-.6*\ipemarkscale) rectangle
      ( .6*\ipemarkscale, .6*\ipemarkscale);
    \coordinate () at (0,0);
  },
  ipe fsquare/.pic={
    \filldraw[line width=0.2*\ipemarkscale, line join=miter]
      (-.5*\ipemarkscale,-.5*\ipemarkscale) rectangle
      ( .5*\ipemarkscale, .5*\ipemarkscale);
    \coordinate () at (0,0);
  },
  ipe cross/.pic={
    \draw[line width=0.2*\ipemarkscale, line cap=butt]
      (-.5*\ipemarkscale,-.5*\ipemarkscale) --
      ( .5*\ipemarkscale, .5*\ipemarkscale)
      (-.5*\ipemarkscale, .5*\ipemarkscale) --
      ( .5*\ipemarkscale,-.5*\ipemarkscale);
    \coordinate () at (0,0);
  },
  /pgf/arrow keys/.cd,
  ipe arrow normal/.style={scale=1},
  ipe arrow tiny/.style={scale=.4},
  ipe arrow small/.style={scale=.7},
  ipe arrow large/.style={scale=1.4},
  ipe arrow normal,
  /tikz/.cd,
  ipe arrows/.style={
    ipe normal/.tip={
      ipe _pointed[length=1bp, width=.666bp, inset=0bp,
                   quick, ipe arrow normal]},
    ipe pointed/.tip={
      ipe _pointed[length=1bp, width=.666bp, inset=0.2bp,
                   quick, ipe arrow normal]},
    ipe linear/.tip={
      ipe _linear[length = 1bp, width=.666bp,
                  ipe arrow normal, quick]},
    ipe fnormal/.tip={ipe normal[fill=white]},
    ipe fpointed/.tip={ipe pointed[fill=white]},
    ipe double/.tip={ipe normal[] ipe normal},
    ipe fdouble/.tip={ipe fnormal[] ipe fnormal},
    ipe arc/.tip={ipe normal},
    ipe farc/.tip={ipe fnormal},
    ipe ptarc/.tip={ipe pointed},
    ipe fptarc/.tip={ipe fpointed},
  },
  ipe arrows, 
]

\tikzset{
  rgb color/.code args={#1=#2}{%
    \definecolor{tempcolor-#1}{rgb}{#2}%
    \tikzset{#1=tempcolor-#1}%
  },
}

\makeatother

\usetikzlibrary{arrows.meta,patterns}
\usetikzlibrary{ipe} 

\begin{document}
\newacro{BN}{bottleneck feature}
\newacro{RTF}{real time factor}
\newacro{DNN}{deep neural network}
\newacro{F0}{fundamental frequency}
\newacro{EER}{equal error rate}
\newacro{WER}{word error rate}
\newacro{PI}{personal information}
\newacro{MSE}{mean-squared error}
\newacro{NSF}{neural source-filter}
\newacro{AM}{acoustic model}
\newacro{VPC}{VoicePrivacy Challenge}
\newacro{ASV}{automated speaker verification}
\newacro{ASR}{automated speech recognition}
\newacro{TDNN}{time delay neural network}
\newacro{FPE}{Fine Pitch Error}
\newacro{GPE}{Gross Pitch Error}
\newacro{MFCC}{Mel Frequency Cepstral Coefficients}
\newacro{VC}{voice conversion}
\newacro{AE}{autoencoder}
\newacro{CNN}{convolutional neural network}
\newacro{PPG}{phonetic posteriorgrams}
\newacro{UAR}{unweighted average recall}
\newacro{PQMF}{pseudo quadrature mirror filterbank}
\newacro{OOD}{out-of-distribution}
\newacro{SER}{speech emotion recognition}
\newacro{OHNN}{orthogonal Householder neural network}
\newacro{GST}{global style tokens}
\newacro{IR}{intermediate representation}
\newacro{VQ}{vector-quantized}
\title{Why disentanglement-based speaker anonymization\\ systems fail at preserving emotions?}

\author{
\IEEEauthorblockN{Ünal Ege Gaznepoglu\IEEEauthorrefmark{1}, Nils Peters\IEEEauthorrefmark{2}}
\IEEEauthorblockA{\IEEEauthorrefmark{1} \textit{International Audio Laboratories Erlangen, Friedrich-Alexander-University Erlangen-Nürnberg, Germany} \\ \thanks{The International Audio Laboratories Erlangen are a joint institution of the Friedrich-Alexander-Universtität Erlangen-Nürnberg and Fraunhofer IIS. Corresponding author: ege.gaznepoglu@audiolabs-erlangen.de}
\IEEEauthorrefmark{2} \textit{Trinity College Dublin, Ireland}}
}
\maketitle
\begin{abstract}
Disentanglement-based speaker anonymization involves decomposing speech into a semantically meaningful representation, altering the speaker embedding, and resynthesizing a waveform using a neural vocoder. State-of-the-art systems of this kind are known to remove emotion information. Possible reasons include mode collapse in GAN-based vocoders, unintended modeling and modification of emotions through speaker embeddings, or excessive sanitization of the \ac{IR}. In this paper, we conduct a comprehensive evaluation of a state-of-the-art speaker anonymization system to understand the underlying causes. We conclude that the main reason is the lack of emotion-related information in the \ac{IR}. The speaker embeddings also have a high impact, if they are learned in a generative context. The vocoder's out-of-distribution performance has a smaller impact. Additionally, we discovered that synthesis artifacts increase spectral kurtosis, biasing emotion recognition evaluation towards classifying utterances as angry. Therefore, we conclude that reporting unweighted average recall alone for emotion recognition performance is suboptimal.
\end{abstract}

\acresetall

\section{Introduction}

Speaker anonymization is the task of modifying a speech signal such that the speaker identity, i.e, timbre, is changed yet other information such as linguistic content and emotions are preserved. Disentanglement-based speaker anonymization consists of obtaining a semantically meaningful speech representation, modifying it to alter the identity, and re-synthesizing a waveform. Numerous systems developed in this spirit have been shown to be successful, in terms of ensuring privacy meanwhile maintaining intelligibility in the main event of this discipline, the \ac{VPC} \cite{champion_vpc_evalplan_2024}.

The \ac{VPC} 2024 evaluation plan \cite{champion_vpc_evalplan_2024} shows that the most successful systems in the literature, such as STTTS (B3) \cite{meyer_prosody_2023}, the neural audio codec-based system (B4) \cite{panariello_speaker_2024}, and the vector quantization-based system (B5) \cite{champion_anonymizing_2024} have a common shortcoming: the loss of emotion information throughout the process, outlined by the gap between emotion recognition performance (UAR, see Sec.~\ref{sec:metrics} for details) on original and anonymized utterances. SSL-SAS, a system that uses self-supervised speech representations, also suffers from this problem \cite{miao_adapting_2024}. Loss of emotion information during speaker anonymization has significant implications for, e.g., medical and legal use cases, where emotional cues may be detrimental for diagnosis or evidence. By understanding the underlying causes, we can develop anonymization systems that preserve emotions, enabling more reliable applications.

Current evaluation methodologies, e.g., the VPC 2024 evaluation plan \cite{champion_vpc_evalplan_2024}, are not specifically designed to identify problematic block(s) of disentanglement-based anonymization systems. In particular, the interplay between the \ac{IR} and the vocoder is challenging to characterize. By performing ablation studies, we can gain further insights into the system design. For example, bypassing the anonymization block has shown that the reconstructions do not necessarily preserve pitch, and contain artifacts \cite{gaznepoglu_evaluation_2023}. The authors of \cite{panariello_vocoder_2023-1} discovered that unified vocoders largely ignore speaker conditioning and cause a phenomenon called vocoder drift, by measuring the discrepancy between the intended pseudo-speaker and the speaker embedding extracted from the output speech. {In a similar spirit, in this paper, we perform a comprehensive ablation study on SSL-SAS, a state-of-the-art speaker anonymization system, to understand why emotion information is lost.}

\section{Hypotheses}

Based on our literature review, we identified three potential culprits that cause emotion information to be lost.

\subsection{Mode collapse of the GAN-based vocoders} \label{hyp:vocoder}

The first possibility is the design and training of the vocoders, which convert the \ac{IR} to a speech waveform. Systems such as \cite{champion_anonymizing_2024} and \cite{miao_adapting_2024} use a joint HiFi-GAN, trained on the reconstruction of read speech (LibriTTS). The adopted methodology (generative adversarial networks) is known to suffer from \textit{mode collapse}, where the model learns to produce a limited subset of possible outputs. The composition of the training dataset, the lack of inductive biases in the system, and the utilized reconstruction loss might hinder the system from generalizing to \ac{OOD} samples, such as expressive speech. Several works \cite{pons_upsampling_2021, bak_avocodo_2023, defossez_high_2023, lee_bigvgan_2023, gu_multi-scale_2024} identified shortcomings and proposed mitigations to improve \ac{OOD} performance.

\subsection{Speaker embeddings capturing emotion data} \label{hyp:spk_emb}

\begin{table*}[t]
    \centering
    \caption{State-of-the-art speaker anonymization systems in the literature, and their speech representations}
    \begin{tabular}{rll}
        \toprule
        \multicolumn{1}{c}{\textbf{System}} & \textbf{Speaker-related representation} & \textbf{Downstream-related representation} \\
        \midrule
        VPC 2024 B3 \cite{meyer_prosody_2023} & GST \cite{wang_style_2018} & Phonetic transcriptions (ling.) + Prosody representation (prosody) \\
        VPC 2024 B4 \cite{panariello_speaker_2024} & Acoustic prompts & Semantic tokens (ling.)\\
        VPC 2024 B5 \cite{champion_anonymizing_2024} & One-hot encoding & Vector-quantized and post-processed wav2vec embeddings (ling.) + F0 (prosody)\\
        SSL-SAS \cite{miao_adapting_2024} & ECAPA-TDNN \cite{desplanques_ecapa-tdnn_2020} & HuBERT-based soft-encoded embeddings (ling.) + F0 (prosody)\\
        \bottomrule
    \end{tabular}
    \label{tab:intermediate_representations} \vspace{-0.5em}
\end{table*}

The penultimate layer of \ac{ASV} systems (e.g., ECAPA-TDNN \cite{desplanques_ecapa-tdnn_2020}) yields a speaker embedding. Speaker anonymization systems then generate a new embedding, called a pseudo-identity, to condition the speech synthesis. Recent work \cite{ulgen_we_2024, ulgen_revealing_2024} has identified that the speaker embeddings obtained in this fashion do not serve well as conditioning information for speech synthesis. Limitations include a tendency to capture undesired emotion information and an indifference to intra-speaker variations. The reported attempts to compensate the speaker embeddings for emotion information were only partially able to bridge the gap between the \ac{UAR} scores of the original and anonymized utterances \cite{miao_adapting_2024}. In particular, the reported recall for sad utterances is still very low, motivating further research.

\subsection{Discretization of the intermediate representation} \label{hyp:ir}

The components of \ac{IR} (see Tab.~\ref{tab:intermediate_representations}) other than the speaker embeddings may be the culprit. We observe that speaker information is removed from the \ac{IR} by discretizing, either via quantization \cite{champion_anonymizing_2024, panariello_speaker_2024}, phonetic transcriptions \cite{meyer_prosody_2023}, or soft-encoding \cite{miao_adapting_2024}. On the contrary, works such as \cite{ghosh_emo-stargan_2023} discovered that indirect emotion supervision, e.g., by controlling emotions through some continuous acoustic features such as spectral centroids, spectral kurtosis, loudness, or $\Delta$F0 is more effective than providing explicit supervision via discrete labels.

{Besides, studies probing HuBERT discovered that earlier transformer layers (up to T3) encode prosodic information \cite{lin_utility_2023}. In contrast, SSL-SAS uses either T6 or T12 as they encode the linguistic content \cite{pasad_comparative_2023}, hinting that even the unsanitized \ac{IR} has a low availability of prosodic information. To summarize, while striving for better privacy, the excessive sanitization might be causing the already scarce emotion information to be lost.}

\section{Methodology}

\subsection{System Under Test}

\begin{figure}[hb]
    \centering
    \includegraphics[width=\linewidth]{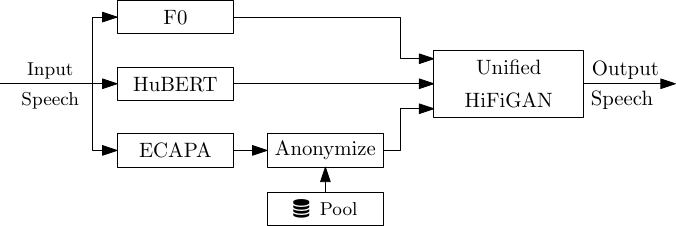}
    \caption{The signal flow diagram of the SSL-SAS system.}
    \label{fig:system}
\end{figure}

\begin{table*}[t]
\vspace{-1em}
    \centering
    \caption{\footnotesize \ac{VPC}2024 evaluation results of our experiments on Libri-test (\acs{EER}, \acs{WER}) and IEMOCAP-test (\acs{WER}, Recall, \acs{UAR}). \textbf{Bold-face} values indicate top-scoring experiments. Results for original data and \ac{VPC}2024 baselines B3-B5 are included for reference.} 
    \begin{tabular}{c c c r r r r r r r r r}
    \toprule
     & & & \multicolumn{3}{c}{Libri-test} & \multicolumn{6}{c}{IEMOCAP-test} \\
    \cmidrule(lr){4-6} \cmidrule(lr){7-12}
     & & & Female & Male &  &                                                                                                     & Sad  & Neutral & Angry & Happy &  All \\
    \cmidrule(lr){4-4} \cmidrule(lr){5-5} \cmidrule(lr){8-8} \cmidrule(lr){9-9} \cmidrule(lr){10-10} \cmidrule(lr){11-11}
     {Speaker emb.} & HuBERT Layer & Post-proc. & \multicolumn{2}{c}{EER [\%] (\ua)} & \multicolumn{2}{c}{WER [\%] (\da)} & \multicolumn{4}{c}{Recall [\%] (\ua)} & \multicolumn{1}{c}{UAR [\%] (\ua)} \\
    \cmidrule(lr){1-1} \cmidrule(lr){2-3}    \cmidrule(lr){4-5} \cmidrule(lr){6-7} \cmidrule(lr){8-11} \cmidrule(lr){12-12}
    \multicolumn{3}{c}{Original data}    &                  8.76 &                   0.42 &              1.84 &             25.39 & 72.58 & 71.66 &  72.82 & 67.19 & 71.06          \\
    \cmidrule(lr){1-1} \cmidrule(lr){2-3}    \cmidrule(lr){4-5} \cmidrule(lr){6-7} \cmidrule(lr){8-11} \cmidrule(lr){12-12}
    \multicolumn{3}{c}{B3}    &                46.53 &                 44.32 &              4.35 &           47.04 & 0.65 & 41.83 &  66.09 & 41.72 & 37.57          \\
    \multicolumn{3}{c}{B4}    &                47.27 &                 49.67 &              5.90 &           50.45 & 11.26 & 46.68 &  61.54 & 54.64 & 42.78          \\
    \multicolumn{3}{c}{B5}    &                49.09 &                 49.22 &              4.37 &             42.99 & 5.07 & 55.30 &  56.20 & 36.10 & 38.17          \\
    \cmidrule(lr){1-1} \cmidrule(lr){2-3}    \cmidrule(lr){4-5} \cmidrule(lr){6-7} \cmidrule(lr){8-11} \cmidrule(lr){12-12}
    N/A            & T1  & N/A            &                 10.76 &                   2.45 &              2.11 &    \textbf{34.95} & \textbf{50.80} & 62.04 &  72.58 & 73.69 & \textbf{64.78} \\
    ECAPA-Orig     & T6  & N/A            &                  5.29 &                   3.56 &              2.15 &    34.98          & 31.58 & 61.68 &  65.67 & 77.54 &          59.11 \\
    ECAPA-Orig     & T12 & soft-label     &                  6.93 &                   4.90 &              2.28 &    37.26          & 12.86 & \textbf{69.20} &  73.50 & 58.44 &          53.50 \\
    GST-Orig       & T12  & soft-label    &                 15.15 &                   9.80 &              2.30 &    37.46          & 29.28 & 75.26 &  60.00 & 48.85   &            53.35 \\
    N/A            & T6  & N/A            &                 11.32 &                   9.55 &              2.31 &    36.25          & 16.40 & 51.79 &  66.79 & \textbf{79.95} &          53.73 \\
    ECAPA-Anon     & T6  & N/A            &                 16.59 &                  11.02 &     \textbf{2.11} &    35.51          & 25.77 & 64.80 &  63.88 & 70.59 &          56.26 \\
    ECAPA-Anon     & T12 & soft-label     &        \textbf{24.27} &         \textbf{27.17} &              2.42 &    38.85          &  3.42 & 46.71 &  \textbf{77.65} & 52.98 &          45.19 \\
    \bottomrule
    \end{tabular}
    \label{tab:results}
    \vspace{-0.5em}
\end{table*}

The system under test, SSL-SAS \cite{miao_language-independent_2022}, is depicted in Fig.~\ref{fig:system}. First, three feature extractors encode distinct components of speech.  ECAPA-TDNN embeddings \cite{desplanques_ecapa-tdnn_2020} encode the speaker identity. The linguistic content is encoded using HuBERT \cite{hsu_hubert_2021}, followed by soft-encoding \cite{van_niekerk_comparison_2022}. For prosody YAAPT \cite{zahorian_spectraltemporal_2008} extracts F0. YAAPT yields a pitch track with a frame rate (per second) of 100, whereas HuBERT outputs have a frame rate of 50, which is upsampled to 100 by repetition as in \cite{miao_language-independent_2022}. Utterance-level speaker embeddings are also repeated. Then, a pseudo-speaker embedding replaces the input speaker embedding. In \cite{miao_language-independent_2023}, the authors of SSL-SAS introduced a novel anonymizer called \ac{OHNN} but since the source code is not released, we use the anonymizer they refer to as \texttt{selec-anon}. First proposed by \cite{fang_speaker_2019}, it uses a speaker pool \add{LibriTTS \texttt{train-other-500}} to select speakers away from the original speaker, then averages a random subset of these candidates to generate a pseudo-identity. Finally, a HiFiGAN-variant, trained on the reconstruction of LibriTTS \texttt{train-clean-100} from the concatenated \ac{IR}, synthesizes the final waveform.

\subsection{The experiments}

{We manipulate the \ac{IR} of SSL-SAS to attribute the loss of emotion information to any of the hypotheses. Table~\ref{tab:results} provides a summary of our experiments. We change the chosen HuBERT transformer layer, and the applied post-processing, to grasp the impact of the downstream-related \ac{IR}. T6 without post-processing and T12 with soft-encoding were the choices in \cite{miao_language-independent_2022}. T1 without post-processing is motivated by \cite{lin_utility_2023}, as it has been shown to encode prosodic information.}
{In addition, we run some experiments to quantify the effect of the speaker embeddings. We substitute ECAPA-TDNN with \ac{GST} to see the impact of speaker embeddings learned in a generative context. Next, we omit them altogether, expecting the vocoder to rely on the information in downstream-related \ac{IR} to reconstruct the utterances. Finally, we use \texttt{selec-anon} to check if anonymizers have any impact on the emotions.}

{In our experiments, we always use the implementation and checkpoints for ECAPA-TDNN and HuBERT (incl. soft-encoder if applicable) available on\footnote{https://github.com/nii-yamagishilab/SSL-SAS, commit 07e51d3}. We use the GST implementation and the checkpoint within VPC 2024 B3 (STTTS), available on\footnote{https://github.com/Voice-Privacy-Challenge/Voice-Privacy-Challenge-2024/, commit 9ae90e1 \label{link:vpc_repo}}. For each configuration, we train the vocoder to 100k steps with same losses and hyperparameters as in \cite{miao_language-independent_2022}.}

\subsection{Evaluation Metrics} \label{sec:metrics}

\begin{figure}[t]
    \centering
    \resizebox{\linewidth}{!}{\begingroup
\renewcommand{\baselinestretch}{1} \endlinechar=-1 \input{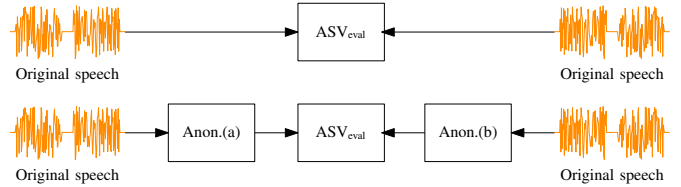}\endgroup \renewcommand{\baselinestretch}{1.5}}
    \caption{Privacy evaluation for speaker anonymization. Top: Unprotected case, for reference. Bottom: The lazy-informed attack model, comparing the resulting speech for two separate invocations of the anonymization system.}
    \label{fig:eval_asv}
\end{figure}
\begin{figure}[t]
    \centering
    \resizebox{\linewidth}{!}{\begingroup
\renewcommand{\baselinestretch}{1} \endlinechar=-1 \input{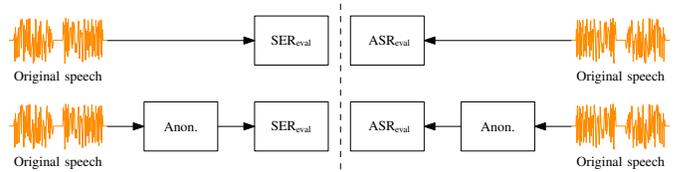}\endgroup \renewcommand{\baselinestretch}{1.5}}
    \caption{Utility evaluation for speaker anonymization. Left: Emotion recognition, Right: Intelligibility via \ac{ASR}.}
    \label{fig:eval_util}
\end{figure}

We use the pretrained part of the VPC 2024 evaluation protocol \cite{champion_vpc_evalplan_2024} \add{as made available on\footnotemark[\value{footnote}]}. To measure privacy benefits, a pretrained $\textrm{ASV}_{eval}$ (see Fig.~\ref{fig:eval_asv}) performs an attack according to attack models of increasing strengths \cite{champion_vpc_evalplan_2024}. VPC 2024 evaluation includes training an $\textrm{ASV}_{eval}^{anon}$ on anonymized speech, to assess the privacy in semi-informed attack scenarios. Since our primary aim is to evaluate the emotion preservation, we instead focus on \ac{ASV} evaluation per the lazy-informed attack model. For any attack model, anonymization is expected to increase the \ac{EER}, hinting that the original speakers of the anonymized utterances could not be identified. 

For utility evaluation (see Fig.~\ref{fig:eval_util}), we use {\ac{VPC}2024} $\textrm{ASR}_{eval}$ to compute \acp{WER}, which are then compared to the performance on the original data. For these experiments, LibriSpeech \cite{panayotov_librispeech_2015} subsets are used. To assess emotion preservation, we use the {\ac{VPC}2024} $\textrm{SER}_{eval}$ along with the IEMOCAP dataset \cite{busso_iemocap_2008}. For both metrics, lower values are desired.

\section{Results}

\subsection{Discretization of the intermediate representation}
We summarize the results of our experiments in Table \ref{tab:results}. The results for SSL-SAS (lowest row, mnemonic \texttt{ECAPA-Anon-T12-soft}) align with \cite{miao_adapting_2024}, and appear more promising than for \ac{VPC} baselines B3-B5 \cite{champion_vpc_evalplan_2024}.  Models that apply quantization to \ac{IR}, such as B5, perform poorer in terms of UAR, compared to SSL-SAS. Switching from T12-soft to T6 causes the most noticeable increase in \ac{UAR}, showing that emotion information is sensitive, more so even than linguistic content, to quantization. Similar behavior is observed for ECAPA-Anon (+11.07 points) as well as ECAPA-Orig (+5.61 points).

Focusing on per-emotion breakdown of \ac{UAR}, we see that there is a huge variation between different emotion classes, unlike the performance on the original data. Overall, $\textrm{SER}_{eval}$ fails to identify sad and neutral utterances once they are processed. Upon checking the confusion matrices, we observe that 49\% of the sad utterances are classified as angry after they are anonymized with SSL-SAS. Furthermore, the recall of angry samples is higher for SSL-SAS (77.65\%) than for the original data (72.82\%), corroborating the observation that anonymization causes the utterances to be perceived as angry by $\textrm{SER}_{eval}$.

\subsection{Speaker embeddings capturing emotion data}
Changing the identity using \texttt{selec-anon} has a smaller impact on the emotion recognition results than the soft-encoding approach (T6: -2.85 points, T12-soft: -8.31). {\texttt{None-T6}, the experiment without any speaker embedding in the \ac{IR}, results in a \ac{UAR} and \ac{WER} worse than \texttt{ECAPA-Orig-T6} (-5.38 points) and \texttt{ECAPA-Anon-T6} (-2.53 points). This shows that HuBERT T6 and F0 alone are not sufficient for high-quality resynthesis. Interestingly, \texttt{None-T6} attains the best recall for happy utterances among all experiments.}

Including the corresponding utterance-level GST embedding yields a similar \ac{UAR} to the ECAPA experiment, showing that speaker embeddings of both generative and discriminative kind contain some emotion information. However, \texttt{GST-Orig-T12-soft} exhibits a slightly more balanced emotion recognition performance than \texttt{ECAPA-Orig-T12-soft}, which attains recall on par with the original data for angry, happy and neutral samples, yet performs far worse for sad utterances (ECAPA: 12.86\% vs. GST: 29.28\%).


\subsection{Mode collapse of the GAN-based vocoders}
Notably, the \ac{WER} results on IEMOCAP-test reveal a significant difference in performance, compared to Libri-test (25\% vs. 1.84\%). This "base effect" is explainable by two factors: 1) domain mismatch, since $\textrm{ASR}_{eval}$ is trained on reading speech; and 2) the presence of interfering speakers and noise in IEMOCAP dataset. However, the additional 10 point difference hints that processing deteriorates the intelligibility of IEMOCAP utterances in a way that does not occur for Libri utterances. Nevertheless, further research is necessary to attribute this difference to underlying reason(s).

Turning to emotion recognition, using a different HuBERT layer, namely, T1 instead of T6 or T12-soft, achieves a UAR of 64.78\%. The recalls of sad and neutral utterances are still lower than that for happy and angry utterances, which are on par with the performance on original data. This shows that the mode collapse hypothesis has a somewhat limited effect on the emotion recognition performance, since HiFiGAN with original training loss as in \cite{kong_hifi-gan_2020} can achieve decent \ac{UAR} and per-class recall, if an \ac{IR} that contains emotion-related information is used. 

\begin{figure*}[t]
    \centering
    \includegraphics[width=\linewidth]{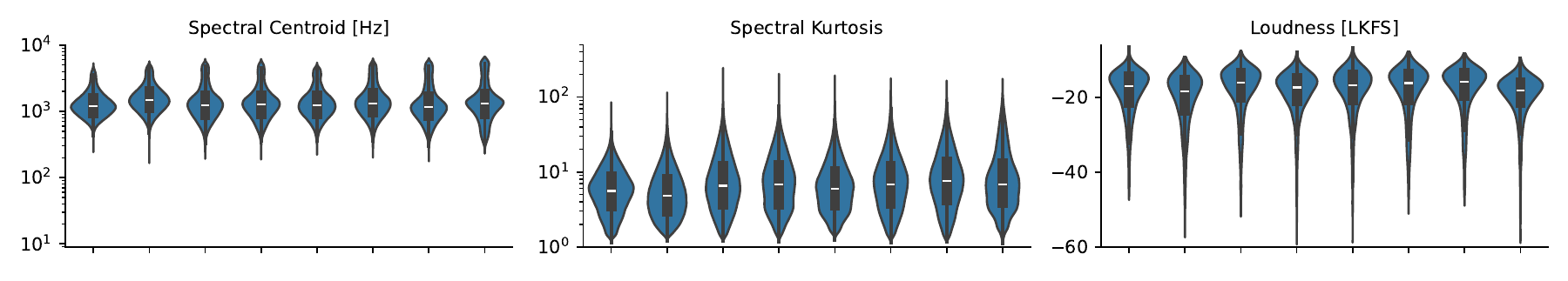}\\
    \vspace{-1em}
    \includegraphics[width=\linewidth]{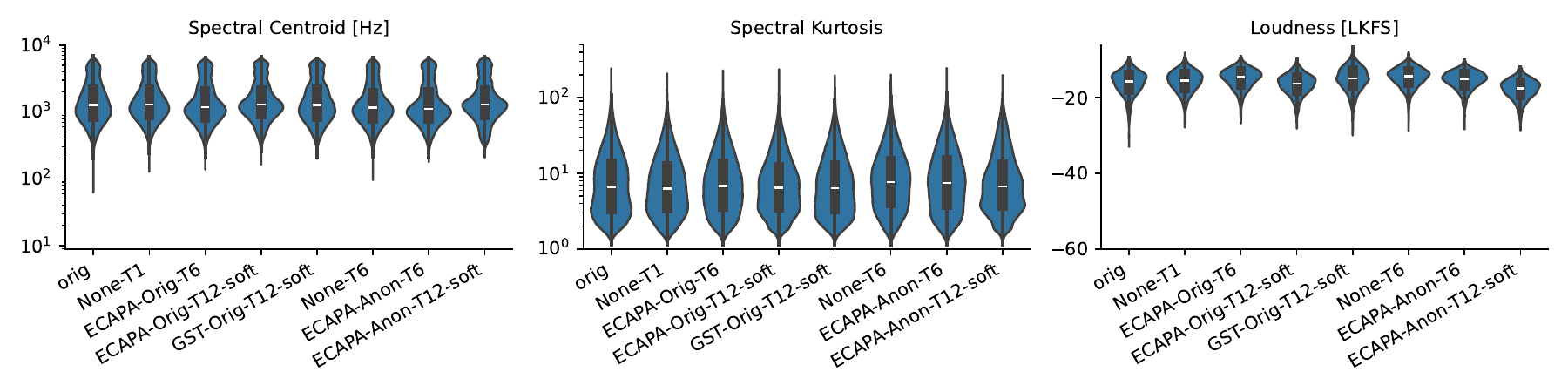}
    \caption{Distributions of emotion-related acoustic features per experiment. Spectral centroid and spectral kurtosis are plotted in log-scale. Top: on IEMOCAP-test, Bottom: on Libri-test}
    \label{fig:distributions}
\end{figure*}

\subsection{What happens to emotion-related acoustic features?}

In this section, we investigate if there is a systematic change in the emotion-related acoustic features (see Sec.~\ref{hyp:ir}). We use the following methodology:

\begin{enumerate}
    \item We identify the voice-active segments of the original IEMOCAP-test and Libri-test samples, using \cite{silero-vad}.
    \item We compute the following acoustic features on the 'oracle' voice active segments for each experiment:
    \begin{itemize}
        \item Spectral Centroid 
        \item Spectral Kurtosis 
        \item Loudness per ITU-R BS.1770-4 as implemented in \cite{noauthor_torchaudiofunctionalloudness_nodate}
    \end{itemize}
\end{enumerate}

Due to space constraints, we only provide an aggregated summary of the acoustic features in Fig.~\ref{fig:distributions}. For randomly selected individual audio samples and their corresponding acoustic feature plots, see our complementary website\footnote{\url{https://audiolabs-erlangen.de/resources/2025-ICASSP-spk-anon-emotion} \label{fn:link}}. 

\subsubsection{Spectral Centroid}
For Libri-test samples, median spectral centroids are found to be approximately 1100 Hz. For all experiments, the higher tail of the distribution moved up, to around 6 kHz, likely due to vocoder artifacts. The experiments except \texttt{ECAPA-Anon-T12-soft}, i.e., SSL-SAS, still have a single modality with comparable median values, whereas for that experiment we see that the distribution has a greater elongation and a second modality is barely identifiable. IEMOCAP samples exhibit a similar pattern, except the lower end of the tail does not go as low as 60 Hz, likely due to increased microphone noise in IEMOCAP utterances.

\subsubsection{Spectral Kurtosis}
This acoustic feature reveals the most interesting insights. For original Libri-test samples, median spectral kurtoses are found to be 6.3. The experiments with mnemonics \texttt{None-T6} and \texttt{ECAPA-Anon} are found to have slightly higher kurtosis values, while the rest have comparable behavior to the original. In contrast, original IEMOCAP-test samples have a lower median spectral kurtosis (5.5), and lower high outliers (80) than the original Libri-test samples. Experiments except \texttt{None-T1} result in a greater medians and greater high outliers. Using HuBERT T12-soft in \ac{IR} introduces non-smooth behavior for both datasets. 

Plotting the spectral kurtosis distributions across different emotion classes, we see that for \texttt{None-T1}, spectral kurtosis medians are ordered from highest to lowest as angry, happy, neutral, and sad. For other experiments, the shapes and the median values become similar and the ordering between medians differ from experiment to experiment. We interpret this as a sign of lost emotion-related acoustic information due to further processing in other experiments. This figure is included in the supplementary material, available on\footnotemark[\value{footnote}]. 

\subsubsection{Loudness}

Original Libri-test samples have a smaller loudness variation than IEMOCAP samples. For both datasets, using HuBERT T12-soft reduces the median loudness. 

\subsection{Limitations of our study}

We have identified 16 samples in IEMOCAP-test, for which the VAD did not yield any segments with speech activity. We noticed that these samples either have a lot of noise, interfering speech, or there is no active speech. A collection of the problematic samples is available on our complementary website\textsuperscript{\ref{fn:link}}. For a more reliable evaluation we suggest the \ac{VPC} organizers to publish a cleaned dataset. Alternatively, researchers could validate on a different dataset such as RAVDESS \cite{livingstone_ryerson_2018}.

In addition, we used \ac{GST} in the way it is introduced in \cite{wang_style_2018}, i.e., trained together with Tacotron, so it learns to encode what is missing from Tacotron's inputs. One could also train or fine-tune \ac{GST} along with SSL-SAS, which should then improve the performance.

\section{Conclusion}

{We have investigated the impact of numerous design choices in speaker anonymization systems on emotion preservation. To grasp the underlying problems, we used a more comprehensive evaluation than the official \ac{VPC} 2024 evaluation protocol, where we gradually reduced the available information in the \ac{IR}. Insufficient emotion information in the \ac{IR} has been identified as the main problem. Quantization, or other approaches such as soft-encoding, limit the emotion information in the \ac{IR}. Moreover, speaker embeddings, especially if learned in a generative context, contain emotion information. The impact of the vocoder (HiFiGAN) is limited compared to the other two hypotheses.}

Based on our results, we think that finding ways to distill emotion information from self-supervised speech representations while preserving the privacy benefits is the main challenge. Compensating or sanitizing the speaker embeddings such that they do not influence the emotion aspect of speech signals, and improving the \ac{OOD} performance of the vocoder is desirable to reach the full potential.

Furthermore, our findings indicate evaluating on a different emotional speech dataset may be more appropriate, as numerous sources of noise and interference are present in IEMOCAP utterances and thus confound the experiment results. In future work, we aim to propose an improved speaker anonymization utilizing the findings of this study.

\balance
\printbibliography[heading=bibnumbered]

@misc{silero-vad,
  author = {{Silero Team}},
  title = {Silero {VAD}: pre-trained enterprise-grade Voice Activity Detector (VAD), Number Detector and Language Classifier},
  year = {2024},
  publisher = {GitHub},
  journal = {GitHub repository},
  howpublished = {\url{https://github.com/snakers4/silero-vad}},
  commit = {insert_some_commit_here},
  email = {hello@silero.ai}
}

@inproceedings{van_niekerk_comparison_2022,
	title = {A Comparison of Discrete and Soft Speech Units for Improved Voice Conversion},
	url = {https://ieeexplore.ieee.org/abstract/document/9746484},
	doi = {10.1109/ICASSP43922.2022.9746484},
	pages = {6562--6566},
	booktitle = {Proc. {IEEE} Intl. Conf. on Acoustics, Speech and Signal Processing ({ICASSP})},
	author = {van Niekerk, Benjamin and Carbonneau, Marc-André and Zaïdi, Julian and Baas, Matthew and Seuté, Hugo and Kamper, Herman},
	urldate = {2025-01-15},
	date = {2022-05},
	note = {{ISSN}: 2379-190X},
}

@article{livingstone_ryerson_2018,
	title = {The Ryerson Audio-Visual Database of Emotional Speech and Song ({RAVDESS}): A dynamic, multimodal set of facial and vocal expressions in North American English},
	volume = {13},
	issn = {1932-6203},
	url = {https://dx.plos.org/10.1371/journal.pone.0196391},
	doi = {10.1371/journal.pone.0196391},
	shorttitle = {The Ryerson Audio-Visual Database of Emotional Speech and Song ({RAVDESS})},
	pages = {e0196391},
	number = {5},
	journaltitle = {{PLOS} {ONE}},
	shortjournal = {{PLoS} {ONE}},
	author = {Livingstone, Steven R. and Russo, Frank A.},
	editor = {Najbauer, Joseph},
	urldate = {2025-01-09},
	date = {2018-05-16},
	langid = {english},
}

@inproceedings{miao_language-independent_2022,
	title = {Language-Independent Speaker Anonymization Approach Using Self-Supervised Pre-Trained Models},
	url = {https://www.isca-archive.org/odyssey_2022/miao22_odyssey.html},
	doi = {10.21437/Odyssey.2022-39},
	eventtitle = {The Speaker and Language Recognition Workshop (Odyssey 2022)},
	pages = {279--286},
	booktitle = {Oddysey: Speaker and Lang. Recognition Workshop},
	author = {Miao, Xiaoxiao and Wang, Xin and Cooper, Erica and Yamagishi, Junichi and Tomashenko, Natalia},
	urldate = {2024-05-29},
	date = {2022-06-28},
	langid = {english},
}

@inproceedings{wang_style_2018,
	title = {Style Tokens: Unsupervised Style Modeling, Control and Transfer in End-to-End Speech Synthesis},
	url = {https://proceedings.mlr.press/v80/wang18h.html},
	shorttitle = {Style Tokens},
	eventtitle = {International Conference on Machine Learning},
	pages = {5180--5189},
	booktitle = {Proc. Intl. Conf. on Machine Learning ({ICML})},
	author = {Wang, Yuxuan and Stanton, Daisy and Zhang, Yu and Ryan, {RJ}-Skerry and Battenberg, Eric and Shor, Joel and Xiao, Ying and Jia, Ye and Ren, Fei and Saurous, Rif A.},
	urldate = {2024-05-27},
	date = {2018-07-03},
	langid = {english},
}

@inproceedings{lin_utility_2023,
	title = {On the Utility of Self-Supervised Models for Prosody-Related Tasks},
	url = {https://ieeexplore.ieee.org/document/10023234/?arnumber=10023234},
	doi = {10.1109/SLT54892.2023.10023234},
	eventtitle = {2022 {IEEE} Spoken Language Technology Workshop ({SLT})},
	pages = {1104--1111},
	booktitle = {Proc. {IEEE} Spoken Lang. Tech. Workshop ({SLT})},
	author = {Lin, Guan-Ting and Feng, Chi-Luen and Huang, Wei-Ping and Tseng, Yuan and Lin, Tzu-Han and Li, Chen-An and Lee, Hung-yi and Ward, Nigel G.},
	urldate = {2024-08-06},
	date = {2023-01},
}

@inproceedings{pasad_comparative_2023,
	title = {Comparative Layer-Wise Analysis of Self-Supervised Speech Models},
	url = {https://ieeexplore.ieee.org/document/10096149},
	doi = {10.1109/ICASSP49357.2023.10096149},
	eventtitle = {{ICASSP} 2023 - 2023 {IEEE} International Conference on Acoustics, Speech and Signal Processing ({ICASSP})},
	pages = {1--5},
	booktitle = {Proc. {IEEE} Intl. Conf. on Acoustics, Speech and Signal Processing ({ICASSP})},
	author = {Pasad, Ankita and Shi, Bowen and Livescu, Karen},
	urldate = {2024-06-24},
	date = {2023-06},
}

@thesis{champion_anonymizing_2024,
	title = {Anonymizing Speech: Evaluating and Designing Speaker Anonymization Techniques},
	url = {http://arxiv.org/abs/2308.04455},
	shorttitle = {Anonymizing Speech},
	institution = {Universite de Lorraine},
	type = {phdthesis},
	author = {Champion, Pierre},
	urldate = {2024-05-27},
	date = {2024-03-01},
	eprinttype = {arxiv},
	eprint = {2308.04455 [cs, eess]},
	note = {00000 },
}

@inproceedings{ulgen_revealing_2024,
	title = {Revealing Emotional Clusters in Speaker Embeddings: A Contrastive Learning Strategy for Speech Emotion Recognition},
	url = {https://ieeexplore.ieee.org/document/10447060},
	doi = {10.1109/ICASSP48485.2024.10447060},
	shorttitle = {Revealing Emotional Clusters in Speaker Embeddings},
	eventtitle = {{ICASSP} 2024 - 2024 {IEEE} International Conference on Acoustics, Speech and Signal Processing ({ICASSP})},
	pages = {12081--12085},
	booktitle = {Proc. {IEEE} Intl. Conf. on Acoustics, Speech and Signal Processing ({ICASSP})},
	author = {Ulgen, Ismail Rasim and Du, Zongyang and Busso, Carlos and Sisman, Berrak},
	urldate = {2024-09-10},
	date = {2024-04},
	note = {{ISSN}: 2379-190X},
}

@online{noauthor_torchaudiofunctionalloudness_nodate,
	title = {torchaudio.functional.loudness — Torchaudio 2.3.0 documentation},
	url = {https://pytorch.org/audio/2.3.0/generated/torchaudio.functional.loudness.html},
	urldate = {2024-09-10},
}

@article{busso_iemocap_2008,
	title = {{IEMOCAP}: interactive emotional dyadic motion capture database},
	volume = {42},
	rights = {http://www.springer.com/tdm},
	issn = {1574-020X, 1574-0218},
	url = {http://link.springer.com/10.1007/s10579-008-9076-6},
	doi = {10.1007/s10579-008-9076-6},
	shorttitle = {{IEMOCAP}},
	pages = {335--359},
	number = {4},
	journaltitle = {Language Resources and Evaluation},
	shortjournal = {Lang Resources \& Evaluation},
	author = {Busso, Carlos and Bulut, Murtaza and Lee, Chi-Chun and Kazemzadeh, Abe and Mower, Emily and Kim, Samuel and Chang, Jeannette N. and Lee, Sungbok and Narayanan, Shrikanth S.},
	urldate = {2024-09-02},
	date = {2008-12},
	langid = {english},
}

@inproceedings{ghosh_emo-stargan_2023,
	title = {Emo-{StarGAN}: A Semi-Supervised Any-to-Many Non-Parallel Emotion-Preserving Voice Conversion},
	url = {http://arxiv.org/abs/2309.07586},
	doi = {10.21437/Interspeech.2023-191},
	shorttitle = {Emo-{StarGAN}},
	pages = {2093--2097},
	booktitle = {Proc. Interspeech Conf.},
	author = {Ghosh, Suhita and Das, Arnab and Sinha, Yamini and Siegert, Ingo and Polzehl, Tim and Stober, Sebastian},
	urldate = {2024-09-01},
	date = {2023-08-20},
	langid = {english},
	eprinttype = {arxiv},
	eprint = {2309.07586 [cs, eess]},
}

@inproceedings{ulgen_we_2024,
	title = {We Need Variations in Speech Synthesis: Sub-center Modelling for Speaker Embeddings},
	url = {http://arxiv.org/abs/2407.04291},
	shorttitle = {We Need Variations in Speech Synthesis},
	booktitle = {{arXiv} preprint},
	publisher = {{arXiv}},
	author = {Ulgen, Ismail Rasim and Busso, Carlos and Hansen, John H. L. and Sisman, Berrak},
	urldate = {2024-07-17},
	date = {2024-07-05},
	langid = {english},
	eprinttype = {arxiv},
	eprint = {2407.04291 [cs, eess]},
	note = {00000 },
}

@inproceedings{gu_multi-scale_2024,
	title = {Multi-Scale Sub-Band Constant-Q Transform Discriminator for High-Fidelity Vocoder},
	url = {https://ieeexplore.ieee.org/document/10448436},
	doi = {10.1109/ICASSP48485.2024.10448436},
	eventtitle = {{ICASSP} 2024 - 2024 {IEEE} International Conference on Acoustics, Speech and Signal Processing ({ICASSP})},
	pages = {10616--10620},
	booktitle = {Proc. {IEEE} Intl. Conf. on Acoustics, Speech and Signal Processing ({ICASSP})},
	author = {Gu, Yicheng and Zhang, Xueyao and Xue, Liumeng and Wu, Zhizheng},
	urldate = {2024-09-01},
	date = {2024-04},
	note = {{ISSN}: 2379-190X},
}

@article{defossez_high_2023,
	title = {High Fidelity Neural Audio Compression},
	journaltitle = {Trans. on Machine Learning Research ({TMLR})},
	author = {Défossez, Alexandre and Copet, Jade and Synnaeve, Gabriel and Adi, Yossi},
	date = {2023},
	langid = {english},
	note = {00395},
}

@inproceedings{lee_bigvgan_2023,
	title = {{BigVGAN}: A Universal Neural Vocoder with Large-Scale Training},
	url = {http://arxiv.org/abs/2206.04658},
	shorttitle = {{BigVGAN}},
	booktitle = {Proc. Intl. Conf. on Learning Representations ({ICLR})},
	publisher = {{arXiv}},
	author = {Lee, Sang-gil and Ping, Wei and Ginsburg, Boris and Catanzaro, Bryan and Yoon, Sungroh},
	urldate = {2024-04-18},
	date = {2023-02-16},
	langid = {english},
	eprinttype = {arxiv},
	eprint = {2206.04658 [cs, eess]},
	note = {00000 },
}

@inproceedings{bak_avocodo_2023,
	title = {Avocodo: Generative Adversarial Network for Artifact-free Vocoder},
	url = {http://arxiv.org/abs/2206.13404},
	shorttitle = {Avocodo},
	booktitle = {Proc. Conf. {AAAI}},
	publisher = {{arXiv}},
	author = {Bak, Taejun and Lee, Junmo and Bae, Hanbin and Yang, Jinhyeok and Bae, Jae-Sung and Joo, Young-Sun},
	urldate = {2023-10-11},
	date = {2023-01-03},
	langid = {english},
	eprinttype = {arxiv},
	eprint = {2206.13404 [cs, eess]},
	note = {00017 },
}

@inproceedings{pons_upsampling_2021,
	location = {Toronto, {ON}, Canada},
	title = {Upsampling Artifacts in Neural Audio Synthesis},
	rights = {https://ieeexplore.ieee.org/Xplorehelp/downloads/license-information/{IEEE}.html},
	isbn = {978-1-72817-605-5},
	url = {https://ieeexplore.ieee.org/document/9414913/},
	doi = {10.1109/ICASSP39728.2021.9414913},
	eventtitle = {{ICASSP} 2021 - 2021 {IEEE} International Conference on Acoustics, Speech and Signal Processing ({ICASSP})},
	pages = {3005--3009},
	booktitle = {Proc. {IEEE} Intl. Conf. on Acoustics, Speech and Signal Processing ({ICASSP})},
	publisher = {{IEEE}},
	author = {Pons, Jordi and Pascual, Santiago and Cengarle, Giulio and Serra, Joan},
	urldate = {2024-04-23},
	date = {2021-06-06},
	langid = {english},
	note = {00000},
}

@inproceedings{miao_adapting_2024,
	title = {Adapting General Disentanglement-Based Speaker Anonymization for Enhanced Emotion Preservation},
	url = {http://arxiv.org/abs/2408.05928},
	booktitle = {{arXiv} preprint},
	publisher = {{arXiv}},
	author = {Miao, Xiaoxiao and Zhang, Yuxiang and Wang, Xin and Tomashenko, Natalia and Soh, Donny Cheng Lock and Mcloughlin, Ian},
	urldate = {2024-09-01},
	date = {2024-08-12},
	langid = {english},
	eprinttype = {arxiv},
	eprint = {2408.05928 [cs, eess]},
}

@inproceedings{panayotov_librispeech_2015,
	title = {Librispeech: An {ASR} corpus based on public domain audio books},
	url = {https://ieeexplore.ieee.org/document/7178964},
	doi = {10.1109/ICASSP.2015.7178964},
	shorttitle = {Librispeech},
	eventtitle = {2015 {IEEE} International Conference on Acoustics, Speech and Signal Processing ({ICASSP})},
	pages = {5206--5210},
	booktitle = {Proc. {IEEE} Intl. Conf. on Acoustics, Speech and Signal Processing ({ICASSP})},
	author = {Panayotov, Vassil and Chen, Guoguo and Povey, Daniel and Khudanpur, Sanjeev},
	urldate = {2024-07-25},
	date = {2015-04},
	note = {06475 
{ISSN}: 2379-190X},
}

@inproceedings{panariello_speaker_2024,
	location = {Seoul, Korea, Republic of},
	title = {Speaker Anonymization Using Neural Audio Codec Language Models},
	rights = {https://doi.org/10.15223/policy-029},
	isbn = {9798350344851},
	url = {https://ieeexplore.ieee.org/document/10447871/},
	doi = {10.1109/ICASSP48485.2024.10447871},
	eventtitle = {{ICASSP} 2024 - 2024 {IEEE} International Conference on Acoustics, Speech and Signal Processing ({ICASSP})},
	pages = {4725--4729},
	booktitle = {Proc. {IEEE} Intl. Conf. on Acoustics, Speech and Signal Processing ({ICASSP})},
	publisher = {{IEEE}},
	author = {Panariello, Michele and Nespoli, Francesco and Todisco, Massimiliano and Evans, Nicholas},
	urldate = {2024-05-21},
	date = {2024-04-14},
	langid = {english},
	note = {00000},
}

@article{hsu_hubert_2021,
	title = {{HuBERT}: Self-Supervised Speech Representation Learning by Masked Prediction of Hidden Units},
	volume = {29},
	rights = {https://ieeexplore.ieee.org/Xplorehelp/downloads/license-information/{IEEE}.html},
	issn = {2329-9290, 2329-9304},
	url = {https://ieeexplore.ieee.org/document/9585401/},
	doi = {10.1109/TASLP.2021.3122291},
	shorttitle = {{HuBERT}},
	pages = {3451--3460},
	journaltitle = {{IEEE}/{ACM} Transactions on Audio, Speech, and Language Processing},
	shortjournal = {{IEEE}/{ACM} Trans. Audio Speech Lang. Process.},
	author = {Hsu, Wei-Ning and Bolte, Benjamin and Tsai, Yao-Hung Hubert and Lakhotia, Kushal and Salakhutdinov, Ruslan and Mohamed, Abdelrahman},
	urldate = {2024-05-30},
	date = {2021},
	langid = {english},
	note = {00000},
}

@article{miao_language-independent_2023,
	title = {Language-independent speaker anonymization using orthogonal Householder neural network},
	journaltitle = {{IEEE}/{ACM} Trans. Audio, Speech, Lang. Process.},
	author = {Miao, Xiaoxiao and Wang, Xin and Cooper, Erica and Yamagishi, Junichi and Tomashenko, Natalia},
	date = {2023-05-30},
	note = {00000},
}

@inproceedings{meyer_prosody_2023,
	title = {Prosody Is Not Identity: A Speaker Anonymization Approach Using Prosody Cloning},
	doi = {10.1109/ICASSP49357.2023.10096607},
	shorttitle = {Prosody Is Not Identity},
	eventtitle = {{ICASSP} 2023 - 2023 {IEEE} International Conference on Acoustics, Speech and Signal Processing ({ICASSP})},
	pages = {1--5},
	booktitle = {Proc. {IEEE} Intl. Conf. on Acoustics, Speech and Signal Processing ({ICASSP})},
	author = {Meyer, Sarina and Lux, Florian and Koch, Julia and Denisov, Pavel and Tilli, Pascal and Vu, Ngoc Thang},
	date = {2023-06},
	note = {00000},
}

@misc{champion_vpc_evalplan_2024,
	title = {3rd {VoicePrivacy} Challenge Evaluation Plan (Version 2.0)},
	author = {{Pierre Champion} and {Nicholas Evans} and {Sarina Meyer} and {Xiaoxiao Miao} and {Michele Panariello} and {Massimiliano Todisco} and {Natalia Tomashenko} and {Emmanuel Vincent} and {Xin Wang}},
	date = {2024},
	langid = {english},
}

@inproceedings{gaznepoglu_evaluation_2023,
	title = {Evaluation of the Speech Resynthesis Capabilities of the {VoicePrivacy} Baseline B1},
	url = {https://www.isca-speech.org/archive/spsc_2023/gaznepoglu23_spsc.html},
	doi = {10.21437/SPSC.2023-10},
	eventtitle = {3rd Symposium on Security and Privacy in Speech Communication},
	pages = {60--64},
	booktitle = {Proc. 3rd Symp. on Security and Privacy in Speech Communication},
	publisher = {{ISCA}},
	author = {Gaznepoglu, Ünal Ege and Peters, Nils},
	urldate = {2023-09-20},
	date = {2023-08-19},
	langid = {english},
	note = {00000},
}

@inproceedings{panariello_vocoder_2023-1,
	title = {Vocoder drift in x-vector-based speaker anonymization},
	booktitle = {Proc. Interspeech Conf.},
	author = {Panariello, Michele and Todisco, Massimiliano and Evans, Nicholas},
	date = {2023-06-05},
	note = {00000},
}

@inproceedings{kong_hifi-gan_2020,
	title = {{HiFi}-{GAN}: Generative Adversarial Networks for Efficient and High Fidelity Speech Synthesis},
	url = {http://arxiv.org/abs/2010.05646},
	doi = {10.48550/arXiv.2010.05646},
	shorttitle = {{HiFi}-{GAN}},
	eventtitle = {Proc. {NeurIPS} Conf.},
	publisher = {{arXiv}},
	author = {Kong, Jungil and Kim, Jaehyeon and Bae, Jaekyoung},
	urldate = {2023-08-01},
	date = {2020-10-23},
	note = {00870},
}

@inproceedings{desplanques_ecapa-tdnn_2020,
	title = {{ECAPA}-{TDNN}: Emphasized Channel Attention, Propagation and Aggregation in {TDNN} Based Speaker Verification},
	url = {http://arxiv.org/abs/2005.07143},
	doi = {10.21437/Interspeech.2020-2650},
	shorttitle = {{ECAPA}-{TDNN}},
	pages = {3830--3834},
	booktitle = {Proc. Interspeech Conf.},
	author = {Desplanques, Brecht and Thienpondt, Jenthe and Demuynck, Kris},
	urldate = {2023-06-13},
	date = {2020-10-25},
	eprinttype = {arxiv},
	eprint = {2005.07143 [cs, eess]},
	note = {00629 },
}

@inproceedings{fang_speaker_2019,
	title = {Speaker anonymization using x-vector and neural waveform models},
	url = {http://arxiv.org/abs/1905.13561},
	booktitle = {Proc. 10th {ISCA} Speech Synthesis Workshop},
	author = {Fang, Fuming and Wang, Xin and Yamagishi, Junichi and Echizen, Isao and Todisco, Massimiliano and Evans, Nicholas and Bonastre, Jean-Francois},
	urldate = {2021-02-18},
	date = {2019-05-29},
	eprinttype = {arxiv},
	eprint = {1905.13561},
	note = {00061 },
}

@article{zahorian_spectraltemporal_2008,
	title = {A spectral/temporal method for robust fundamental frequency tracking},
	volume = {123},
	issn = {1520-8524},
	doi = {10.1121/1.2916590},
	pages = {4559--4571},
	number = {6},
	journaltitle = {The Journal Acoust. Soc. of America},
	shortjournal = {J Acoust Soc Am},
	author = {Zahorian, Stephen A. and Hu, Hongbing},
	date = {2008-06},
	pmid = {18537404},
	note = {00162 },
}

\end{document}